\documentstyle[12pt]{article}

\newcommand{\tr}{\mbox{Tr} }
\newcommand{\ket}[1]{\left | #1 \right \rangle}
\newcommand{\bra}[1]{\left \langle #1 \right |}
\newcommand{\amp}[2]{\left \langle #1 \left | #2 \right. \right \rangle}

\newcommand{\proj}[1]{\ket{#1} \! \bra{#1}}

\newcommand{\superop}{\mbox{\$}}

\newcommand{\hilbert}[1]{{\cal H}_{#1}}

\begin{document}

\begin{center}

\bigskip
{\Large \bf Quantum data processing \\
and error correction}\\
\bigskip
{\large Benjamin Schumacher}$^{(1),(2),(3)}$ \medskip \\
{\large M. A. Nielsen}$^{(1)}$
\bigskip
{\small{\sl

$^{(1)}$Center for Advanced Studies, Department of Physics and Astronomy, \\
University of New Mexico, Albuquerque, NM 87131 \vspace{2mm} \\

$^{(2)}$Theoretical Astrophysics, T-6, MS B288,
Los Alamos National Laboratory, \\ Los Alamos, NM  87545 \vspace{2mm} \\

$^{(3)}$Permanent address: Department of Physics, Kenyon College, 
Gambier, OH  43022
}}

\end{center}

\subsection*{\centering Abstract}
{This paper investigates properties of noisy quantum information
channels.  We define a new quantity called {\em coherent information} 
which measures the amount of quantum information conveyed in the noisy
channel.  This quantity can never be increased by quantum information
processing, and it yields a simple necessary and sufficient condition for
the existence of perfect quantum error correction.}

\vfill

PACS numbers: 03.65.Bz, 89.70.+c\vfill

\newpage

\section{Introduction}

This paper reports some results relating to the transmission of
quantum information through noisy channels, that is, channels which
are not isolated from their environments.  It builds upon an earlier
investigation of this situation by one of us \cite{qfano}.
We begin with a brief general
discussion of noisy quantum processes and their mathematical
descriptions.

Suppose a quantum system $Q$ is subjected to a dynamical evolution,
which may represent the transmission of $Q$ via a noisy quantum channel.
In general, the evolution of $Q$ will be represented by a 
{\em superoperator} $\superop^{Q}$, which gives the mapping from
initial states (represented by density operators $\rho^{Q}$) to final
states:
\begin{displaymath}
	\rho^{Q'} = \superop^{Q} \left ( \rho^{Q} \right )	
\end{displaymath}
where we use primes to denote states after the evolution.  The mapping
represented by $\superop^{Q}$ is linear in $\rho^{Q}$ and preserves both
the trace and the positivity of its arguments.  The evolution of $Q$ will
be unitary only if it is isolated from other systems.  We might represent
this by the following schematic diagram:

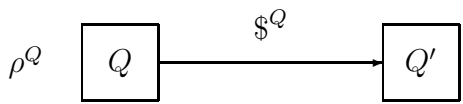
\begin{center}
\unitlength 1cm
\begin{picture}(8,3)(0,2)
\put(2,3){\framebox(1,1){$Q$}}
\put(3,3.5){\vector(1,0){3}}
\put(6,3){\framebox(1,1){$Q'$}}
\put(1.25,3.5){\makebox(0,0){$\rho^{Q}$}}
\put(4.5,4){\makebox(0,0){$\superop^{Q}$}}
\end{picture}
\end{center}

We might imagine, however, that the system $Q$ is part of a larger
system $RQ$, and that this compound system is initially in a pure state
$\ket{\Psi^{RQ}}$.  Then $\rho^{Q} = \tr_{R} \, \proj{\Psi^{RQ}}$.
(We say that $\ket{\Psi^{RQ}}$ is a ``purification'' of $\rho^{Q}$.)
The system $R$ is isolated and has a zero internal Hamiltonian.  
This situation might be
represented by a slightly more complicated diagram:

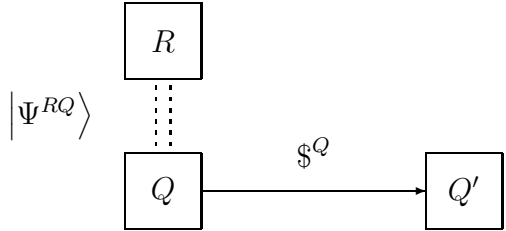
\begin{center}
\unitlength 1cm
\begin{picture}(8,5)(0,2)
\put(2,3){\framebox(1,1){$Q$}}
\put(3,3.5){\vector(1,0){3}}
\put(6,3){\framebox(1,1){$Q'$}}
%
\put(2,5){\framebox(1,1){$R$}}
\put(2.4,4.10){\line(0,1){.05}}	\put(2.6,4.10){\line(0,1){.05}}
\put(2.4,4.25){\line(0,1){.05}}  \put(2.6,4.25){\line(0,1){.05}}
\put(2.4,4.40){\line(0,1){.05}}  \put(2.6,4.40){\line(0,1){.05}}
\put(2.4,4.55){\line(0,1){.05}}  \put(2.6,4.55){\line(0,1){.05}}
\put(2.4,4.70){\line(0,1){.05}}  \put(2.6,4.70){\line(0,1){.05}}
\put(2.4,4.85){\line(0,1){.05}}  \put(2.6,4.85){\line(0,1){.05}}
\put(1,4.5){\makebox(0,0){$\ket{\Psi^{RQ}}$}}
%
\put(4.5,4){\makebox(0,0){$\superop^{Q}$}}
\end{picture}
\end{center}

We will call $R$ the ``reference'' system.  $Q$ may in fact initially be
in a pure entangled state with some external system, but from our point
of view $R$ is introduced simply as a mathematical device 
to purify the initial state.  

The overall system $RQ$ evolves according to the ``extended'' superoperator 
$I^{R} \otimes \superop^{Q}$, where $I^{R}$ is the identity.  That
is, 
\begin{displaymath}
	\rho^{RQ'}  = I^{R} \otimes \superop^{Q} \, 
			\left ( \rho^{RQ} \right ) .
\end{displaymath}
The extended superoperator $I^{R} \otimes \superop^{Q}$ is also 
trace-preserving, and if it is to be a legitimate quantum evolution
it had also better preserve the positivity of the density operator.
This second condition is in fact a non-trivial requirement on the
original superoperator $\superop^{Q}$ called {\em complete positivity}
\cite{cpmaps}.
The physical requirement that $Q$-dynamics be extensible in this 
trivial way to dynamics for the compound system $RQ$ imposes the 
mathematical requirement that the superoperator $\superop^{Q}$ be
completely positive.

It turns out that every completely positive $\superop^{Q}$ has a
representation as a {\em unitary} evolution on a larger system
\cite{repthms}.
That is, if $\superop^{Q}$ is an allowable quantum evolution for $Q$
(one that is extensible as indicated above), then we can introduce an
``environment'' system $E$, initially in a pure state, such that
\begin{displaymath}
	\superop^{Q} (\rho^{Q}) = \tr_{E} \, U^{QE}
		\left ( \rho^{Q} \otimes \proj{0^{E}} \right )
		{U^{QE}}^{\dagger}
\end{displaymath}
for a unitary operator $U^{QE}$.  This might be represented as

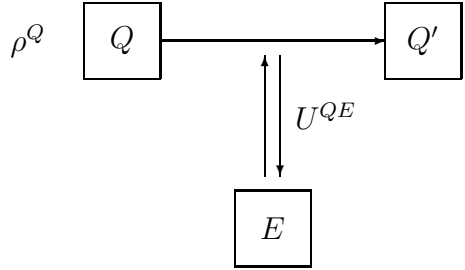
\begin{center}
\unitlength 1cm
\begin{picture}(8,5)(0,0)
\put(2,3){\framebox(1,1){$Q$}}
\put(3,3.5){\vector(1,0){3}}
\put(6,3){\framebox(1,1){$Q'$}}
%
\put(1.25,3.5){\makebox(0,0){$\rho^{Q}$}}
%
\put(4,0.5){\framebox(1,1){$E$}}
\put(4.4,1.7){\vector(0,1){1.6}}
\put(4.6,3.3){\vector(0,-1){1.6}}
\put(4.8,2.5){\makebox(0,0)[cl]{$U^{QE}$}}
\end{picture}
\end{center}

The environment system $E$ and operator $U^{QE}$ might be chosen 
to be the actual physical environment and its interaction with $Q$, 
but this is not really necessary.  The only thing that matters 
for our purposes is the dynamics of $Q$, and $E$ is introduced as
only as a mathematical artifice.  There are many choices of $E$,
$\ket{0^{E}}$, and $U^{QE}$ which will do the same job.  We will call
a representation of $\superop^{Q}$ in terms of a unitary evolution of 
a larger system (with the environment $E$ initially in a pure state)
a {\em unitary representation} of $\superop^{Q}$.  Every allowable
$\superop^{Q}$ admits such a representation.

If we introduce both the reference system $R$ to purify the initial state
and the environment $E$ to give a unitary representation for $\superop^{Q}$,
the situation looks like this:
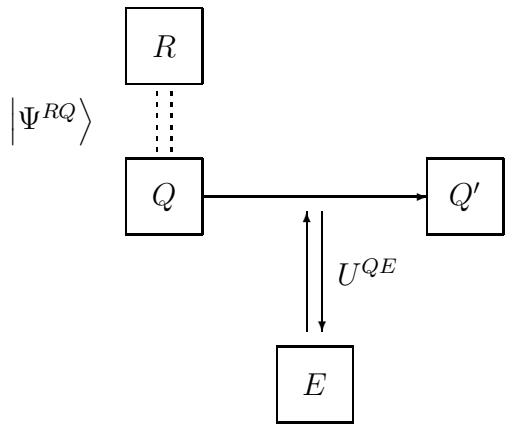
\begin{center}
\unitlength 1cm
\begin{picture}(8,7)(0,0)
\put(2,3){\framebox(1,1){$Q$}}
\put(3,3.5){\vector(1,0){3}}
\put(6,3){\framebox(1,1){$Q'$}}
%
\put(2,5){\framebox(1,1){$R$}}
\put(2.4,4.10){\line(0,1){.05}}	\put(2.6,4.10){\line(0,1){.05}}
\put(2.4,4.25){\line(0,1){.05}}  \put(2.6,4.25){\line(0,1){.05}}
\put(2.4,4.40){\line(0,1){.05}}  \put(2.6,4.40){\line(0,1){.05}}
\put(2.4,4.55){\line(0,1){.05}}  \put(2.6,4.55){\line(0,1){.05}}
\put(2.4,4.70){\line(0,1){.05}}  \put(2.6,4.70){\line(0,1){.05}}
\put(2.4,4.85){\line(0,1){.05}}  \put(2.6,4.85){\line(0,1){.05}}
\put(1,4.5){\makebox(0,0){$\ket{\Psi^{RQ}}$}}
%
\put(4,0.5){\framebox(1,1){$E$}}
\put(4.4,1.7){\vector(0,1){1.6}}
\put(4.6,3.3){\vector(0,-1){1.6}}
\put(4.8,2.5){\makebox(0,0)[cl]{$U^{QE}$}}
\end{picture}
\end{center}
The initial pure state of the joint system $RQE$ is 
\begin{displaymath}
	\ket{\Psi^{RQE}} = \ket{\Psi^{RQ}} \otimes \ket{0^{E}} .
\end{displaymath}
Since the overall evolution is unitary, the final state is also a pure
state:
\begin{displaymath}
	\ket{\Psi^{RQE'}} =  \left ( 1^{R} \otimes U^{QE} \right )
				\ket{\Psi^{RQE}} .
\end{displaymath}
The states of the various subsystems before and after the evolution
may be obtained from these states by partial traces.

It is also possible to represent $\superop^{Q}$ in an ``intrinsic'' way,
one that does not introduce any additional quantum systems.
One particularly useful representation of this sort
is the {\em operator-sum representation}, which involves a collection
of operators $A^{Q}_{\mu}$ that act in the Hilbert space $\hilbert{Q}$
describing $Q$.  This is
\begin{displaymath}
	\rho^{Q'} =
	\superop^{Q} (\rho^{Q}) = \sum_{\mu} A^{Q}_{\mu} \rho^{Q}
				{A^{Q}_{\mu}}^{\dagger}.
\end{displaymath}
The operators $A^{Q}_{\mu}$ must satisfy a normalization condition:
\begin{displaymath}
	\sum_{\mu} {A^{Q}_{\mu}}^{\dagger} A^{Q}_{\mu}  =  1^{Q} .
\end{displaymath}
If we have an operator-sum representation for $\superop^{Q}$, then 
we can easily write down an operator-sum representation for the
extended superoperator $I^{R} \otimes \superop^{Q}$ using 
the operators $1^{R} \otimes A^{Q}_{\mu}$.

The following three conditions are equivalent \cite{repthms}:
\begin{itemize}
	\item  $\superop^{Q}$ is a trace-preserving, completely positive
		linear map on density operators of $Q$.
	\item  $\superop^{Q}$ has a unitary representation.
	\item  $\superop^{Q}$ has a normalized operator-sum representation.
\end{itemize}
For a given $\superop^{Q}$, neither the unitary representation nor the
operator-sum representation is unique.

\section{Entanglement fidelity}

From now on, we will suppose that
the system $Q$, initially in the state $\rho^{Q}$, is subjected to the 
evolution operator $\superop^{Q}$.  We may introduce a reference system
$R$ to purify the initial state to $\ket{\Psi^{RQ}}$, and we may introduce
a unitary representation for $\superop^{Q}$ involving an environment
system $E$, as convenient.  Nevertheless, our focus will be on quantities 
that are {\em intrinsic} to $Q$, depending only on $\rho^{Q}$ and 
$\superop^{Q}$.

Given a pure state $\ket{\psi}$ of a quantum system, we can define the 
{\em fidelity} $F$ of an arbitrary (possibly mixed) state $\rho$ of 
the system as 
\begin{displaymath}
	F = \bra{\psi} \rho \ket{\psi} .
\end{displaymath}
$F$ is a measure of ``how close'' $\rho$ is to $\proj{\psi}$, and is 
equal to unity if and only if $\rho = \proj{\psi}$.  (It is possible
to extend the definition of fidelity to a measure of closeness between
two arbitrary density operators $\rho_{1}$ and $\rho_{2}$, but this 
simpler definition is sufficient for our purposes \cite{qfidelity}.)

The first important intrinsic property of $Q$ we will define
is the {\em entanglement fidelity} $F_{e}$.  This is \cite{qfano}
\begin{eqnarray}
	F_{e} 	& = &	\bra{\Psi^{RQ}} \rho^{RQ'} \ket{\Psi^{RQ}} 
			\nonumber \\
		& = &	\sum_{\mu} \left ( \tr \rho^{Q} A^{Q}_{\mu} \right )
			\left ( \tr \rho^{Q} {A^{Q}_{\mu}}^{\dagger} \right ).
\end{eqnarray}
According to the first expression, $F_{e}$ measures how faithfully the
entangled state $\ket{\Psi^{RQ}}$ is preserved by the dynamics of $Q$.
The second expression emphasizes that this is a quantity intrinsic to 
$Q$, i.e., depending only on $\rho^{Q}$ and $\superop^{Q}$.  The exact
way that $\rho^{Q}$ is ``purified'' into $\ket{\Psi^{RQ}}$ is irrelevant.

It is useful to explore the relation between $F_{e}$ and various other
fidelities that may be defined for $Q$.

Suppose we have an ensemble ${\cal E}$ of pure states, in which 
the $i$th state $\ket{\psi^{Q}_{i}}$ occurs with probability $p_{i}$.  
The ensemble is described by the density operator
\begin{displaymath}
	\rho^{Q} = \sum_{i} p_{i} \proj{\psi^{Q}_{i}} .
\end{displaymath}
If we subject
the $i$th state to the dynamical superoperator $\superop^{Q}$, the resulting
state is $\rho^{Q'}_{i} = \superop^{Q} \left ( \proj{\psi^{Q}_{i}} 
\right)$.  The ``input-output'' fidelity of this process is 
\begin{displaymath}
	F_{i} = \bra{\psi^{Q}_{i}} \rho^{Q'}_{i} \ket{\psi^{Q}_{i}} .
\end{displaymath}  
The average fidelity $\bar{F}$ for the ensemble ${\cal E}$ is given by
\begin{displaymath}
	\bar{F} = \sum_{i} p_{i} F_{i}
	= \sum_{i} p_{i} \bra{\psi^{Q}_{i}} \rho^{Q'}_{i} \ket{\psi^{Q}_{i}} .
\end{displaymath}

Given $\rho^{Q}$ and $\superop^{Q}$ we can also define the entanglement
fidelity $F_{e}$.  It turns out that this entanglement 
fidelity is never greater than the average fidelity \cite{qfano}:  
\begin{equation}
	F_{e} \leq \bar{F}.  
\end{equation}
Thus, the entanglement fidelity is a lower bound for the average fidelity
of an ensemble of pure states.

We will briefly sketch the reasons for this connection between 
$F_{e}$ and $\bar{F}$.
Given a purification $\ket{\Psi^{RQ}}$ for
$\rho^{Q}$, we can always realize the ensemble
${\cal E}$ as an ensemble of relative states of $Q$ given by the outcomes
of the measurement of an observable on $R$.  In other words, the entangled
state of $RQ$ allows us to create the ensemble ${\cal E}$ of $Q$ states by
a procedure that affects only $R$.  This procedure commutes with the 
dynamics of the system $Q$ given by $\superop^{Q}$, and so could be 
performed after $Q$ has undergone its dynamical evolution.  
This allows us to express $\bar{F}$ as the probability of a measurement
outcome on the evolved state
$\rho^{RQ'}$, and this condition turns out to be weaker than the 
condition that expresses $F_{e}$.  Thus $F_{e} \leq \bar{F}$.

Now, given any state $\ket{\phi^{Q}}$ in the subspace that supports
$\rho^{Q}$, it is always possible to find an ensemble 
${\cal E}$ for $\rho^{Q}$
in which $\ket{\phi^{Q}}$ is a component with non-vanishing probability.
This has an interesting implication.  Let 
\begin{displaymath}
	F = \bra{\phi^{Q}} \superop^{Q} \left ( 
		\proj{\phi^{Q}} \right ) \ket{\phi^{Q}}
\end{displaymath}
be the ``input-output'' fidelity associated with $\ket{\phi^{Q}}$.
Then $F_{e} = 1$ only if $F = 1$ for all states $\ket{\phi^{Q}}$ in the  
support of $\rho^{Q}$.  This is because can find an ensemble
for $\rho^{Q}$ containing $\ket{\phi^{Q}}$, and the average fidelity
$\bar{F}$ of that ensemble must be unity.  It follows that the fidelity
of every component of the ensemble is unity.

Another connection between the pure state fidelity and the entanglement
fidelity is this:  Let $\eta \geq  0$ and suppose 
$F \geq 1 - \eta$ for all $\ket{\phi^{Q}}$ in the support of $\rho^{Q}$.
Then it can be shown that \cite{laf96} $F_{e} \geq 1 - (3 \eta/2)$
(a similar result was also pointed out to us by Howard Barnum
\cite{barnum}).
Thus we can conclude
that $F_{e} = 1$ if and only if $F = 1$ for every pure state $\ket{\psi^{Q}}$
in the support of $\rho^{Q}$.

The entanglement fidelity $F_{e}$, which depends only on 
$\rho^{Q}$ and $\superop^{Q}$, thus has some useful relations to the 
other fidelities of the system $Q$. We might informally summarize these
by saying that a high entanglement fidelity $F_{e}$
implies a high ensemble average fidelity $\bar{F}$, 
and a high minimum fidelity on the supporting subspace of 
$\rho^{Q}$ implies that the entanglement
fidelity $F_{e}$ cannot be too much lower.

\section{Entropy production}

The second important intrinsic quantity that we will define
is the {\em entropy production} $S_{e}$ \cite{qfano}.
Let $S(\rho) = - \tr \rho \log \rho$ be the von Neumann entropy of a 
density operator $\rho$ (where the logarithm is take to be base 2).  Then
\begin{eqnarray*}
	S_{e}	& = &	S(\rho^{RQ'})  =  S(\rho^{E'})	\\
		& = &	S(W)
\end{eqnarray*}
where $W$ is a density operator with components (in an orthonormal basis)
\begin{displaymath}
	W_{\mu \nu} =  \tr A^{Q}_{\mu} \rho^{Q} {A^{Q}_{\nu}}^{\dagger} .
\end{displaymath}
Once again, $S_{e}$ has an easy interpretation in terms of the entangled
state $\ket{\Psi^{RQ}}$, as the entropy of the joint system $RQ$ after the
evolution (or, equivalently, the entropy of the environment $E$ afterwards if
the environment starts out in a pure state).  Nevertheless, $S_{e}$ is an
intrinsic property of $Q$, depending only on $\rho^{Q}$ and $\superop^{Q}$.

The entropy production is not in general equal to the changes in entropy
either of the system $Q$ or the actual physical environment of $Q$.
It is a measure of the information exchanged between $Q$ and the rest
of the world during the evolution $\superop^{Q}$.  It has several useful
properties; for example, it limits the amount of information that an
eavesdropper might acquire in a quantum cryptographic protocol \cite{qfano}.

A connection between $F_{e}$ and $S_{e}$ is given by the {\em quantum Fano
inequality} \cite{qfano}, which states that, 
if the Hilbert space $\hilbert{Q}$ describing
system $Q$ has $d$ complex dimensions,
\begin{displaymath}
	h(F_{e}) + (1-F_{e}) \log (d^{2} - 1)  \geq  S_{e}
\end{displaymath}
where $h(p) = - p \log p - (1-p) \log (1-p)$.  This means, among other things,
that if $F_{e} = 1$ then $S_{e} = 0$.  (The quantum Fano inequality is 
analogous to the classical Fano inequality \cite{cover}, which gives a roughly
similar relation between the probability of error in a classical channel
and an entropy term describing the noise in the channel.)

\section{Coherent quantum information}

We now define a third intrinsic quantity of interest, which we will call
{\em coherent (quantum) information} $I_{e}$.  This may be defined as
\begin{eqnarray}
	I_{e}	& = &	S(\rho^{Q'}) - S(\rho^{RQ'})	\nonumber \\
		& = &	S(\rho^{Q'}) - S_{e} .
\end{eqnarray}
(This obviously depends only on $\rho^{Q}$ and $\superop^{Q}$.)  $I_{e}$
may be positive, negative, or zero.  An analogous quantity for
classical systems can never be positive, since the entropy of the joint
system $RQ$ can never be less than the entropy of the subsystem $Q$.  Thus,
we can think of $I_{e}$ as measuring the ``non-classicity'' of the final 
joint state $\rho^{RQ'}$, the degree of quantum entanglement retained by
$R$ and $Q$.

Phrased in this way, $I_{e}$ is a natural measure of the degree
to which quantum coherence is retained by the dynamical process 
$\superop^{Q}$.

We will begin exploring the properties of $I_{e}$ by making use of the
{\em subadditivity} of the von Neumann entropy \cite{wehrl}.
Consider a compound system $AB$ composed of subsystems $A$ and $B$.  Then
\begin{displaymath}
	S(\rho^{AB}) \leq S(\rho^{A}) + S(\rho^{B}) .
\end{displaymath}
Equality holds if and only if $AB$ is in a product state $\rho^{AB} = 
\rho^{A} \otimes \rho^{B}$.  

A second useful fact applies if $AB$ is in a pure state.
In this case $\rho^{A}$ and $\rho^{B}$ have exactly the same non-zero 
eigenvalues (as can be easily seen from the Schmidt decomposition of
$\ket{\Phi^{AB}}$, and thus $S(\rho^{A}) = S(\rho^{B})$.

Now suppose that we have a unitary representation for $\superop^{Q}$ 
involving an environment system $E$ initially in a pure state.  The total
system $RQE$ is thus initially in a pure state; and since it evolves 
according to unitary dynamics, the final state of $RQE$ is also pure.
The coherent information $I_{e}$ is thus
\begin{eqnarray*}
	I_{e} & = & S(\rho^{Q'}) - S_{e} 	\\
	      & = & S(\rho^{RE'}) - S(\rho^{E'}) \\
	      & \leq & S(\rho^{R'}) ,
\end{eqnarray*}
where the last inequality follows by subadditivity.  But $R$ is not affected
by the interaction between $Q$ and $E$, so $\rho^{R'} = \rho^{R}$.  Since
$RQ$ is initially in a pure state, we conclude that
\begin{equation}
	S(\rho^{Q})  \geq  \underbrace{S(\rho^{Q'}) - S_{e}}_{I_{e}} .
\end{equation}
The coherent information can be no greater than the initial entropy
of $Q$, which measures the initial degree of entanglement of $R$ and $Q$.
($S(\rho^{Q})$ also measures the resources necessary to faithfully
store this entanglement.)
Equality holds if and only if $\rho^{RE'} = \rho^{R} \otimes \rho^{E'}$.

This is a special case of a more general property of the coherent 
information $I_{e}$, which we will demonstrate in the next section.

\section{Quantum data processing inequality}

Suppose $X$, $Y$, and $Z$ are classical random variables, and suppose that
\begin{displaymath}
	X \longrightarrow Y \longrightarrow Z
\end{displaymath}
is a Markov process, so that $Z$ depends only on $Y$ and not on $X$ directly.
For example, $X$ and $Y$ might be the input and output of a noisy 
communication channel, and $Z$ might be the result of some (possibly
stochastic) processing of the output.  It is possible to prove a ``data
processing inequality'' \cite{cover}
for classical information theory, which states that
\begin{displaymath}
	I(X:Z) \leq I(X:Y)
\end{displaymath}
where $I(X:Z)$ is the mutual information between $X$ and $Z$, etc.  This
means that the mutual information between the input and output of a channel
cannot be increased by processing the output in any way.

We can establish a similar inequality for the coherent information $I_{e}$.
Suppose the initial state of $Q$ is $\rho^{Q}$ (which has a purification
$\ket{\Psi^{RQ}}$), and further suppose that $Q$ undergoes two successive
dynamical evolutions, described by superoperators $\superop^{Q}_{1}$ and
$\superop^{Q}_{2}$.  Then
\begin{eqnarray*}
	\rho^{Q'}  & = &  \superop^{Q}_{1}(\rho^{Q})	\\
	\rho^{Q''} & = &  \superop^{Q}_{2}(\rho^{Q'})	\\
		   & = &  \superop^{Q}_{2} \circ \superop^{Q}_{1} (\rho^{Q}) .
\end{eqnarray*}
We will call the evolution by $\superop^{Q}_{1}$ the ``first stage'' of
the evolution, and the evolution by $\superop^{Q}_{2}$ the ``second stage''.
These might represent, for example, the transmission of the 
information in $Q$ through a noisy channel (described by $\superop^{Q}_{1}$)
followed by some quantum information processing such as error correction
(described by $\superop^{Q}_{2}$.)  Our schematic is:
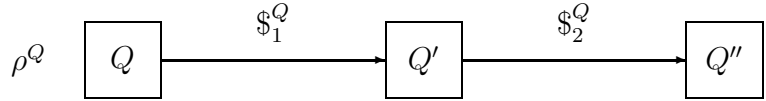
\begin{center}
\unitlength 1cm
\begin{picture}(12,3)(0,2)
\put(2,3){\framebox(1,1){$Q$}}
\put(3,3.5){\vector(1,0){3}}
\put(6,3){\framebox(1,1){$Q'$}}
%
\put(7,3.5){\vector(1,0){3}}
\put(10,3){\framebox(1,1){$Q''$}}
\put(1.25,3.5){\makebox(0,0){$\rho^{Q}$}}
%
\put(4.5,4){\makebox(0,0){$\superop^{Q}_{1}$}}
\put(8.5,4){\makebox(0,0){$\superop^{Q}_{2}$}}
\end{picture}
\end{center}
The overall process is represented by the composition of these two
processes, so that $\superop^{Q}_{12} = \superop^{Q}_{2} \circ
\superop^{Q}_{1}$.

We adopt adopt unitary representations for these processes.  That is, 
we imagine that there are two environment systems $E_{1}$ and $E_{2}$, 
initially in pure states $\ket{0^{E_{1}}}$ and $\ket{0^{E_{2}}}$, which 
interact in succession with $Q$ via unitary operators $U^{QE_{1}}$ and 
$V^{QE_{2}}$.  The full schematic diagram, including the reference system 
$R$, looks like
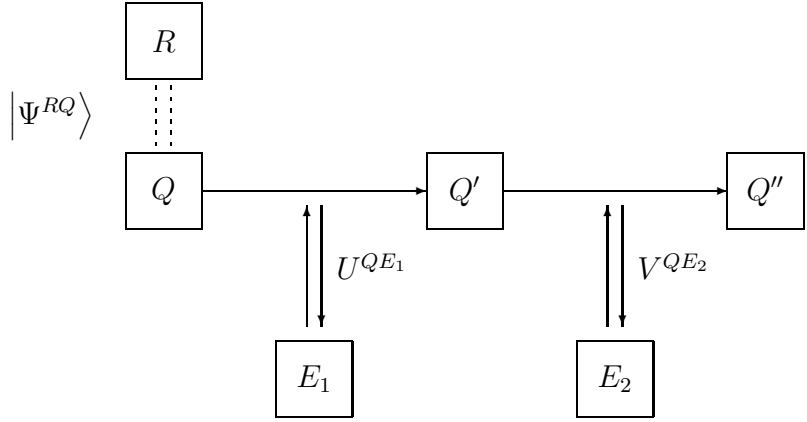
\begin{center}
\unitlength 1cm
\begin{picture}(12,7)(0,0)
\put(2,3){\framebox(1,1){$Q$}}
\put(3,3.5){\vector(1,0){3}}
\put(6,3){\framebox(1,1){$Q'$}}
%
\put(7,3.5){\vector(1,0){3}}
\put(10,3){\framebox(1,1){$Q''$}}
%
\put(2,5){\framebox(1,1){$R$}}
\put(2.4,4.10){\line(0,1){.05}}	\put(2.6,4.10){\line(0,1){.05}}
\put(2.4,4.25){\line(0,1){.05}}  \put(2.6,4.25){\line(0,1){.05}}
\put(2.4,4.40){\line(0,1){.05}}  \put(2.6,4.40){\line(0,1){.05}}
\put(2.4,4.55){\line(0,1){.05}}  \put(2.6,4.55){\line(0,1){.05}}
\put(2.4,4.70){\line(0,1){.05}}  \put(2.6,4.70){\line(0,1){.05}}
\put(2.4,4.85){\line(0,1){.05}}  \put(2.6,4.85){\line(0,1){.05}}
\put(1,4.5){\makebox(0,0){$\ket{\Psi^{RQ}}$}}
%
%
\put(4,0.5){\framebox(1,1){$E_{1}$}}
\put(4.4,1.7){\vector(0,1){1.6}}
\put(4.6,3.3){\vector(0,-1){1.6}}
\put(4.8,2.5){\makebox(0,0)[cl]{$U^{QE_{1}}$}}
%
\put(8,0.5){\framebox(1,1){$E_{2}$}}
\put(8.4,1.7){\vector(0,1){1.6}}
\put(8.6,3.3){\vector(0,-1){1.6}}
\put(8.8,2.5){\makebox(0,0)[cl]{$V^{QE_{2}}$}}
\end{picture}
\end{center}
The initial state of the whole system is
\begin{displaymath}
	\ket{\Psi^{RQE_{1}E_{2}}} = \ket{\Psi^{RQ}} \otimes \ket{0^{E_{1}}}
						\otimes \ket{0^{E_{2}}} .
\end{displaymath}
In the first stage of the dynamics, this evolves to 
\begin{eqnarray*}
    \ket{\Psi^{RQE_{1}{E_{2}}'}}
	& = & 	\left ( 1^{R} \otimes U^{QE_{1}} \otimes 1^{E_{2}} \right )
		\ket{\Psi^{RQE_{1}E_{2}}}  \\
	& = &	\ket{\Psi^{RQ{E_{1}}'}} \otimes \ket{0^{E_{2}}} .
\end{eqnarray*}
In the second stage, this evolves to 
\begin{displaymath}
\ket{\Psi^{RQE_{1}{E_{2}}''}}
	 = \left ( 1^{R} \otimes 1^{E_{1}} \otimes V^{QE_{2}} \right )
		\ket{\Psi^{RQE_{1}{E_{2}}'}}  .
\end{displaymath}
The states of the subsystems can be derived by partial traces of these.

To analyze this two-stage process, we make use of a property of the von
Neumann entropy called {\em strong subadditivity} \cite{wehrl}.
Let $ABC$ be a compound system composed of three subsystems
$A$, $B$, and $C$.  Then
\begin{displaymath}
	S(\rho^{ABC}) + S(\rho^{B})  \leq  S(\rho^{AB}) + S(\rho^{BC}) .
\end{displaymath}
This property is logically stronger than simple subadditivity;
if $B$ is supposed to be in a pure state (so that $\rho^{ABC} = 
\rho^{AB} \otimes \proj{\phi^{B}}$), then we recover ordinary subaddtivity
for $A$ and $C$.

We will apply this inequality to the compound system $RE_{1}E_{2}$ after
both stages of the dynamics have taken place.  This yields
\begin{displaymath}
	S(\rho^{RE_{1}{E_{2}}''}) + S(\rho^{{E_{1}}''})
		\leq  S(\rho^{R{E_{1}}''})  +  S(\rho^{E_{1}{E_{2}}''}).
\end{displaymath}
Each term in this inequality may be re-written in a different form.  
For example, since the overall state of $RQE_{1}E_{2}$ is pure at every
stage, it follows that
\begin{displaymath}
	S(\rho^{RE_{1}{E_{2}}''}) = S(\rho^{Q''}) .
\end{displaymath}
Neither of the systems $R$ or $E_{1}$ are involved in the second stage
of the dynamics, in which $Q$ and $E_{2}$ interact.  Thus, their state
does not change during this stage:  $\rho^{R{E_{1}}''} = \rho^{R{E_{1}}'}$.  
After the first stage, as noted above, the compound system $RQE_{1}$ 
is in a pure state.  Thus,
\begin{displaymath}
	S(\rho^{R{E_{1}}''}) = S(\rho^{R{E_{1}}'}) = S(\rho^{Q'}).
\end{displaymath}
The remaining two terms can both be recognized as entropy productions
of various processes.  That is,
\begin{eqnarray*}
	S(\rho^{{E_{1}}''}) & = & S(\rho^{{E_{1}}''})  =  S_{e1}	\\
	S(\rho^{E_{1}{E_{2}}''}) & = & S_{e12} 
\end{eqnarray*}
where $S_{e1}$ is the entropy production of the first stage, and $S_{e12}$
is the overall entropy production of both stages.  Note that 
in general $S_{e12} \neq S_{e1} + S_{e2}$.  In fact, the overall entropy
production $S_{e12}$ can be less than either of the individual entropy 
productions $S_{e1}$ and $S_{e2}$.

Making these substitutions,
the strong subaddtivity inequality for $RE_{1}E_{2}$ after both stages of
the dynamics yields
\begin{eqnarray*}
	S(\rho^{Q''}) + S_{e1} &  \leq  &  S(\rho^{Q'}) + S_{e12} \\
	S(\rho^{Q''}) - S_{e12} & \leq  &  S(\rho^{Q'}) - S_{e1} .
\end{eqnarray*}
That is, $I_{e12} \leq I_{e1}$.  The coherent information in the first stage
cannot be increased by the additional dynamics of the second stage.  
We thus can summarize our results so far as follows:
\begin{equation}
	S(\rho^{Q}) 
	\geq \underbrace{S(\rho^{Q'}) - S_{e1}}_{I_{e1}} 
	\geq \underbrace{S(\rho^{Q''}) - S_{e12}}_{I_{e12}}.
\end{equation}
This is the quantum data processing inequality.  (The first inequality, of
course, is a special case of the second, since $S(\rho^{Q})$ is the 
coherent information in the trivial process given by $\superop^{Q} = I^{Q}$.)

\section{Error correction}

Suppose $\superop^{Q}_{1}$ represents the transmission of quantum information
via a noisy channel.  $\superop^{Q}_{1}$ may involve ``decoherence'' and 
other noise processes, which will reduce the entanglement fidelity $F_{e1}$
of the channel.  However, it has been shown that under some circumstances
it is possible to do {\em quantum error correction} on the output of the
channel, restoring the initial state of the system either exactly or very
nearly by an allowable quantum process \cite{qerror}.  
This error correction process 
typically consists of an incomplete measurement performed on $Q$ 
followed by a unitary evolution of $Q$ that depends on the measurement 
outcome.  We will describe our quantum error correction scheme by the evolution
superoperator $\superop^{Q}_{2}$, so the overall process of 
channel-dynamics-plus-error-correction is given by 
$\superop^{Q}_{12} = \superop^{Q}_{2} \circ \superop^{Q}_{1}$.

The question naturally arises, under what circumstances can quantum error
correction be performed?  We will consider an interesting special case of
this question:  Given some channel dynamics $\superop^{Q}_{1}$, when is
it possible to find a subsequent quantum evolution $\superop^{Q}_{2}$ which
gives {\em perfect} error correction?

We will take perfect error correction to mean that the entanglement fidelity
$F_{e12}$ of the overall process is unity.  In other words, we require that
the error correction scheme be able to perfectly restore the entanglement of
$Q$ with the system $R$.  (This is a reasonable definition, since we know
that the entanglement fidelity equals unity if and only if every pure state
in the subspace supporting $\rho^{Q}$ has fidelity unity.)
If $F_{e} = 1$ then the final (mixed) state of $Q$ must equal
the initial state:  $\rho^{Q''} = \rho^{Q}$.  From the quantum Fano inequality
we can also infer that the entropy production $S_{e12}$ of the overall
process must be zero.

The quantum data processing inequality allows us to establish
a necessary condition for the existence of a perfect error correction scheme.
If $\superop^{Q}_{2}$ is such a scheme for the initial state $\rho^{Q}$ and
the channel dynamics $\superop^{Q}_{1}$, then
\begin{eqnarray*}
   S(\rho^{Q})	& \geq &	S(\rho^{Q'}) - S_{e1}	\\
		& \geq &	S(\rho^{Q''}) - S_{e12}	\\
		& = &		S(\rho^{Q})		\\
   S(\rho^{Q})	& = &		S(\rho^{Q'}) - S_{e1}  = I_{e1} .
\end{eqnarray*}
Thus, perfect error correction is possible only if the coherent information
of the channel equals the entropy of the input state.

We will next show that $S(\rho^{Q}) = I_{e1}$ is also a {\em sufficient} 
condition for the existence of a perfect error correction scheme.

We begin by writing down the Schmidt decomposition of the initial 
pure entangled state $\ket{\Psi^{RQ}}$ of the system $RQ$:
\begin{displaymath}
	\ket{\Psi^{RQ}} = \sum_{k} \sqrt{\lambda_{k}} \,
			\ket{\alpha^{R}_{k}} \otimes \ket{\beta^{Q}_{k}} 
\end{displaymath}
where we take the sum to include all of the non-zero eigenvalues
$\lambda_{k}$ of $\rho^{Q}$ (and thus also $\rho^{R}$).
If $S(\rho^{Q}) = I_{e1}$, then we have already shown that
$\rho^{R{E_{1}}'} = \rho^{R} \otimes \rho^{{E_{1}}'}$.  This means that
\begin{displaymath}
	\rho^{R{E_{1}}'} = \sum_{kl} \lambda_{k} \mu_{l} 
			\proj{\alpha^{R}_{k}} \otimes \proj{\gamma^{E_{1}}_{l}}
\end{displaymath}
where the $\mu_{l}$ are the non-zero eigenvalues of $\rho^{{E_{1}}'}$ and
$\ket{\gamma^{E_{1}}_{l}}$ are the corresponding eigenstates.

The overall state of $RQE_{1}$ is a pure state $\ket{\Psi^{RQ{E_{1}}'}}$.  We
can use our expression for $\rho^{R{E_{1}}'}$ to write down a Schmidt 
decomposition of this overall state (separating into subsystems $Q$ and 
$RE_{1}$):
\begin{displaymath}
	\ket{\Psi^{RQ{E_{1}}'}} = \sum_{kl} \sqrt{\lambda_{k} \mu_{l}} \,
		\ket{\alpha^{R}_{k}} \otimes \ket{\phi^{Q}_{kl}}
			\otimes \ket{\gamma^{E_{1}}_{l}} .
\end{displaymath}
Here the $Q$ states $\ket{\phi^{Q}_{kl}}$ are orthonormal and span a
subspace of $\hilbert{Q}$.  Let $\Pi^{Q}$ be the projection onto the subspace
perpendicular to this one, so that
\begin{displaymath}
	\Pi^{Q} +  \sum_{kl} \proj{\phi^{Q}_{kl}}  =  1^{Q} .
\end{displaymath}

We will now explicitly construct an operator-sum representation for the
error correction process $\superop^{Q}_{2}$, and show that it is a perfect
error-correction scheme.  Let
\begin{eqnarray*}
	A^{Q}_{0} & = & \Pi^{Q} \\
	A^{Q}_{l} & = & \sum_{k} \ket{\beta^{Q}_{k}} \bra{\phi^{Q}_{kl}} .
\end{eqnarray*}
Intuitively, for each $l$ the operator $A^{Q}_{l}$ represents a projection
onto the subspace spanned by the vectors $\ket{\phi^{Q}_{kl}}$ (for all 
values of $k$), followed by a unitary transformation that takes 
$\ket{\phi^{Q}_{kl}}$ to $\ket{\beta^{Q}_{k}}$.
It is easy to see that
\begin{eqnarray*}
	A^{Q}_{0} \ket{\phi^{Q}_{kl}} & = & 0	\\
	A^{Q}_{l} \ket{\phi^{Q}_{kl'}} & = & \delta_{ll'} 
		\, \ket{\beta^{Q}_{k}} .
\end{eqnarray*}
To yield an allowable dynamical evolution of $Q$, these must be properly
normalized.  That is, 
\begin{eqnarray*}
	{A^{Q}_{0}}^{\dagger} A^{Q}_{0}  
         +  \sum_{l} {A^{Q}_{l}}^{\dagger} A^{Q}_{l} & = &
	\Pi^{Q} + \sum_{l} \sum_{k k'} \ket{\phi^{Q}_{kl}} 
		\amp{\beta^{Q}_{k}}{\beta^{Q}_{k'}} \bra{\phi^{Q}_{k'l}} \\
	& = & \Pi^{Q} + \sum_{kl} \proj{\phi^{Q}_{kl}}  =  1^{Q} .
\end{eqnarray*}
Thus, the operators $A^{Q}_{0}$ and $A^{Q}_{l}$ yield an operator-sum
representation of an allowed quantum evolution superoperator 
$\superop^{Q}_{2}$.

To see that $\superop^{Q}_{2}$ specifies a perfect error-correction scheme,
consider the effect on $\ket{\Psi^{RQ{E_{1}}'}}$ 
of the ``extended'' superoperator 
$I^{R} \otimes \superop^{Q} \otimes I^{E_{1}}$.
The operator-sum representation of 
$I^{R} \otimes \superop^{Q} \otimes I^{E_{1}}$ 
is composed of operators of the form 
$1^{R} \otimes A^{Q}_{l} \otimes 1^{E_{1}}$.
\begin{eqnarray*}
	1^{R} \otimes A^{Q}_{0} \otimes 1^{E_{1}} \ket{\Psi^{RQ{E_{1}}'}}
		& = & \sum_{kl} \sqrt{\lambda_{k} \mu_{l}} \,
			\ket{\alpha^{R}_{k}} \otimes
			\left ( A^{Q}_{0} \ket{\phi^{Q}_{kl}} \right )
			\otimes \ket{\gamma^{E_{1}}_{l}}		\\
	& = & 0 \\
	1^{R} \otimes A^{Q}_{l} \otimes 1^{E_{1}} \ket{\Psi^{RQ{E_{1}}'}}
		& = & \sum_{kl'} \sqrt{\lambda_{k} \mu_{l'}} \,
			\ket{\alpha^{R}_{k}} \otimes
			\left ( A^{Q}_{l} \ket{\phi^{Q}_{kl'}} \right )
			\otimes \ket{\gamma^{E_{1}}_{l'}}		\\
		& = & \sum_{k} \sqrt{\lambda_{k} \mu_{l}} \,
			\ket{\alpha^{R}_{k}} \otimes 
			\ket{\beta^{Q}_{k}} \otimes \ket{\gamma^{E_{1}}_{l}} \\
		& = & \sqrt{\mu_{l}} \left ( 
			\sum_{k} \sqrt{\lambda_{k}} \, \ket{\alpha^{R}_{k}} 
			\otimes \ket{\beta^{Q}_{k}} \right ) \otimes
			\ket{\gamma^{E}_{l}}				\\
		& = & \sqrt{\mu_{l}} \, \ket{\Psi^{RQ}} \otimes 
			\ket{\gamma^{E}_{l}} .
\end{eqnarray*}
Therefore,
\begin{eqnarray*}
   \rho^{RQ{E_{1}}''}
	& = &	I^{R} \otimes \superop^{Q}_{2} \otimes I^{E_{1}}
		\left ( \proj{\Psi^{RQ{E_{1}}'}} \right )	\\
	& = &	\left ( 1^{R} \otimes A^{Q}_{0} \otimes 1^{E_{1}} \right )
		\proj{\Psi^{RQ{E_{1}}'}} 
		\left ( 1^{R} \otimes A^{Q}_{0} \otimes 1^{E_{1}} 
		\right )^{\dagger} \\
	&   &	+ \sum_{l} \left (
		1^{R} \otimes A^{Q}_{l} \otimes 1^{E_{1}} \right )
		\proj{\Psi^{RQ{E_{1}}'}}
		\left (1^{R} \otimes A^{Q}_{l} \otimes 1^{E_{1}}
		\right )^{\dagger}	\\
	& = &	\sum_{l} \mu_{l} \, \, \proj{\Psi^{RQ}} 
		\otimes \proj{\gamma^{E_{1}}_{l}} \\
	& = &	\proj{\Psi^{RQ}} \otimes \rho^{{E_{1}}'} .
\end{eqnarray*}
The final state of $RQ$ is $\rho^{RQ''} = \tr_{E_{1}} \rho^{RQ{E_{1}}''}
= \proj{\Psi^{RQ}}$, which is exactly the original entangled state. 
Therefore the entanglement fidelity of the entire
process is $F_{e12} = 1$.  Our superoperator $\superop^{Q}$ thus gives
a perfect error correction scheme.

Once again we emphasize that, although we made use of the particular input
state $\rho^{Q}$ (with purification $\ket{\Psi^{RQ}}$) to construct our
perfect error-correction scheme, this is equivalent to perfect error
correction for all pure states in the support of $\rho^{Q}$, or indeed for
any other entangled state with the same support in $\hilbert{Q}$.

We may compare our result to a classical theorem \cite{cover}.  
Suppose the random
variables $X$ and $Y$ represent the inputs and outputs of a classical
information channel.  Then the input may be reconstructed from the output
with zero probability of error if and only if $H(X) = I(X:Y)$, where
$H(X)$ is the Shannon entropy of the input variable.

\section{Remarks}

The condition $S(\rho^{Q}) = I_{e}$, which is necessary and sufficient for
perfect quantum error correction, has some interesting implications, which
we will briefly mention here.

Suppose that the state $\rho^{Q}$ is due to an ensemble ${\cal E}$ in
which the pure state $\ket{\psi^{Q}_{i}}$ appears with probability 
$p_{i}$.  As we remarked before, we can realize such an ensemble by
starting with a purification $\ket{\Psi^{RQ}}$ and performing a measurement
of a suitable $R$ observable.  The $i$th outcome of this measurement 
will appear with probability $p_{i}$ and the relative state of $Q$ 
given that measurement will be $\ket{\psi^{Q}_{i}}$.  The question
``Which Q state?'' is then equivalent to the question ``Which $R$
measurement outcome?''

As we showed, the equality $S(\rho^{Q}) = I_{e}$
means that $\rho^{RE'} = \rho^{R} \otimes 
\rho^{E'}$.  If $R$ and $E$ are in a product state after the dynamical
evolution, then measurement results on $E$ have no statistical correlation
with measurement results on $R$.  In other words,
no observable on $E$ alone will be able to provide any information about
the outcome of a measurement performed on $R$.  Therefore, no $E$ observable
can provide any information about which $Q$ state from the ensemble ${\cal E}$
is present.  In short, perfect quantum error correction is possible only if
the environment obtains no information about the state of the system $Q$.

We summarize our main points and conclusions here.
\begin{itemize}
	\item  For a given initial state $\rho^{Q}$ and dynamical 
		superoperator $\superop^{Q}$, we may define several
		intrinsic quantities of interest, including the
		entanglement fidelity $F_{e}$, the entropy production
		$S_{e}$, and the coherent information $I_{e}$.  $F_{e}$
		and $S_{e}$ are related by a quantum version of the
		Fano inequality of classical information theory.
	\item  The entanglement fidelity is closely related to various
		``input-output'' fidelities for pure states of $Q$.
	\item  The coherent information is a measure of the amount of
		``distinctively quantum'' information that passes
		through a channel.  In general, $S(\rho^{Q}) \geq I_{e}$.
	\item  The coherent information can never be increased by 
		the action of further dynamics, so that for successive
		independent processes 1 and 2 we obtain the quantum
		data processing inequality, $I_{e1} \geq I_{e12}$.
	\item  Perfect quantum error correction is possible if and only
		if $S(\rho^{Q}) = I_{e}$, in which case the environment
		has obtained no information about the state of $Q$ via
		its interaction with $Q$.
\end{itemize}

In general, we believe that the coherent information $I_{e}$ will play
a role in quantum information theory analogous to that played by the
mutual information $I(X:Y)$ in the classical theory.  There are many
differences between the two.  For one thing, the coherent information
has a built-in ``time asymmetry'', being defined for an input state
$\rho^{Q}$ and a process $\superop^{Q}$, while the mutual information
$I(X:Y)$ is a symmetric quantity built out of a joint probability
distribution for $X$ and $Y$ in which time does not explicitly appear.
For another,
$I_{e} = S(\rho^{Q'}) - S(\rho^{RQ'})$ is a quantity that can never be
positive classically, so that no classical channel can convey a positive
amount of coherent quantum information.

The paper continues the program of finding useful ``intrinsic'' quantities 
by ``extrinsic'' means, introducing a reference system $R$ to purify the
initial state and an environment $E$ to make the overall dynamics unitary.
This approach appears to yield many important new insights into quantum 
information theory.

\section*{Acknowledgments}

The authors wish to acknowledge the help of many people during the course
of this work, including
H. Barnum, C. H. Bennett, C. M. Caves, C. A. Fuchs, E. H. Knill, 
R. Laflamme, J. Smolin, M. D. Westmoreland, and W. H. Zurek. This work
was supported in part by the Office of Naval Research
(Grant No. N00014-93-1-0116). M. A. Nielsen
acknowledges the support of a Fulbright Scholarship.

\end{document}